\documentclass[final,1p,times]{elsarticle_mod}
\usepackage[english]{babel}
\usepackage[latin1]{inputenc} 

\usepackage{fancyhdr}
\usepackage{indentfirst} 

\usepackage{graphicx}  
\usepackage{subfigure}

\usepackage{bm}
\usepackage{amssymb} 
\usepackage{amsmath}
\usepackage{latexsym}
\usepackage{amsthm}
\usepackage{hyperref}
\usepackage{lineno}

\begin{document}
\begin{frontmatter}

\title{Measurement of the atmospheric $\nu_\mu$ energy spectrum from 100 GeV to 200 TeV with the ANTARES telescope}
\begin{abstract}

Atmospheric neutrinos are produced during cascades initiated by the interaction of primary cosmic rays with air nuclei. In this paper, a measurement of the atmospheric $\nu_\mu + \bar{\nu}_\mu$ energy spectrum in the energy range $0.1 - 200$ TeV is presented, using data collected by the ANTARES underwater neutrino telescope from 2008 to 2011. Overall, the measured flux is $\sim$25\% higher than predicted by the conventional neutrino flux, and compatible with the measurements reported in ice.
The flux is compatible with a single power-law dependence with spectral index $\gamma_{meas}=3.58\pm 0.12$. With the present statistics the contribution of prompt neutrinos cannot be established.

\end{abstract}

\author[UPV]{S.~Adri\'an-Mart\'inez}
\author[Colmar]{A. Albert}
\author[CPPM]{I. Al Samarai}
\author[UPC]{M.~Andr\'e}
\author[Genova]{M. Anghinolfi}
\author[Erlangen]{G. Anton}
\author[IRFU/SEDI]{S. Anvar}
\author[UPV]{M. Ardid}
\author[NIKHEF]{T.~Astraatmadja\fnref{tag:1}}
\author[CPPM]{J-J. Aubert}
% ...
\newpage

\address[UPV]{\scriptsize{Institut d'Investigaci\'o per a la Gesti\'o Integrada de les Zones Costaneres (IGIC) - Universitat Polit\`ecnica de Val\`encia. C/  Paranimf 1 , 46730 Gandia, Spain.}}
\address[Colmar]{\scriptsize{GRPHE - Institut universitaire de technologie de Colmar, 34 rue du Grillenbreit BP 50568 - 68008 Colmar, France }}
\address[CPPM]{\scriptsize{CPPM, Aix-Marseille Universit\'e, CNRS/IN2P3, Marseille, France}}
\address[UPC]{\scriptsize{Technical University of Catalonia, Laboratory of Applied Bioacoustics, Rambla Exposici\'o,08800 Vilanova i la Geltr\'u,Barcelona, Spain}}
\address[Genova]{\scriptsize{INFN - Sezione di Genova, Via Dodecaneso 33, 16146 Genova, Italy}}
\address[Erlangen]{\scriptsize{Friedrich-Alexander-Universit\"at Erlangen-N\"urnberg, Erlangen Centre for Astroparticle Physics, Erwin-Rommel-Str. 1, 91058 Erlangen, Germany}}
\address[IRFU/SEDI]{\scriptsize{Direction des Sciences de la Mati\`ere - Institut de recherche sur les lois fondamentales de l'Univers - Service d'Electronique des D\'etecteurs et d'Informatique, CEA Saclay, 91191 Gif-sur-Yvette Cedex, France}}
\address[NIKHEF]{\scriptsize{Nikhef, Science Park,  Amsterdam, The Netherlands}}
% ...
\address[IRFU/SPP]{\scriptsize{Direction des Sciences de la Mati\`ere - Institut de recherche sur les lois fondamentales de l'Univers - Service de Physique des Particules, CEA Saclay, 91191 Gif-sur-Yvette Cedex, France}}
% ...
\address[Genova-UNI]{\scriptsize{Dipartimento di Fisica dell'Universit\`a, Via Dodecaneso 33, 16146 Genova, Italy}}
% ...

\fntext[tag:1]{\scriptsize{Also at University of Leiden, the Netherlands}}
\fntext[tag:2]{\scriptsize{On leave of absence at the Humboldt-Universit\"at zu Berlin}}
\fntext[tag:3]{\scriptsize{Also at Accademia Navale de Livorno, Livorno, Italy}}
\fntext[tag:4]{\scriptsize{Now at the Max Planck Institute for Physics, Munich}}
\cortext[ca]{Corresponding authors, email: lfusco@bo.infn.it}

\end{frontmatter}

\section{Introduction} \label{intro}

Cosmic neutrinos propagate without significant losses from very distant sources, and so isotropic diffuse flux generated by the ensemble of all cosmic sources in the Universe is expected. 
The flux of atmospheric neutrinos represents an irreducible background for neutrino astronomy, including the search for a diffuse flux, and must be subtracted from the expected cosmic signal. 
Due to the low fluxes and the extremely small neutrino cross-sections, 
neutrino telescopes require very large instrumented volumes.
Muon neutrinos and antineutrinos that undergo charged current weak interactions in the vicinity of the instrumented volume produce detectable muons. 
In the following the $\nu_\mu$ and $\overline \nu_\mu$ are referred as  \textit{muon neutrinos}, or $\nu_\mu$. 

Atmospheric muons and neutrinos are produced in the showers of high energy cosmic rays in the Earth's atmosphere. Below 100 GeV, the $\nu_\mu$ flux as a function of the zenith angle for different event topologies is modulated by neutrino oscillations, as measured by the SuperKamiokande \cite{sk-mu}, MACRO \cite{macro-mu} and Soudan 2 \cite{soudan} experiments.
Recently, neutrino telescope data were used to measure the oscillation parameters of atmospheric neutrinos using muon tracks induced by atmospheric neutrinos with energies as low as 20 GeV \cite{ANTAosci,ICosci}.
Increasingly detailed calculations of the atmospheric neutrino flux have appeared in the last decade as uncertainties on their flux become a limiting factor for fundamental physics studies using atmospheric neutrinos (neutrino mass and mixing, mass hierarchy \cite{smirnov}). 
The flux of atmospheric neutrinos in the TeV (or higher) energy range is extrapolated from lower energies and from knowledge of the primary cosmic ray flux and mass composition. 

Standard neutrino mixing (as described by the $3\times 3$-dimensional Pontecorvo-Maki-Nakagawa-Sakata matrix will not modify the atmospheric $\mu_\nu$ flux above $\sim$0.1 TeV. Thus, accurate measurements of the atmospheric flux allow the investigation of non-standard effects - such as Lorentz invariance violation, of the equivalence principle or other new physics effects (see, for example, Ref. \cite{lore2} and references therein). For instance the MACRO experiment used high ($>$ 130 GeV) energy neutrinos to bound the Lorentz invariance violating parameters \cite{battistoni}.

The $\nu_\mu$  energy spectrum up to 1 TeV was measured by the Frejus collaboration \cite{frejus} and derived from SuperKamiokande data \cite{Gonzales}. 
The energy spectrum of atmospheric $\nu_\mu$ in the hundreds of TeV region has been obtained by the AMANDA \cite{amanda_ff,amanda_atm} and IceCube \cite{ic_atm} Collaborations, with values differing up to 50\%, although compatible within their respective uncertainties.
The ability of neutrino telescopes to measure the incoming neutrino direction and energy is particularly relevant for the measurement presented here. The ice properties and the efficiency of the photomultiplier tubes are the dominant contribution to systematic uncertainties. 
The differences between under ice and under water neutrino telescopes concerning the reconstruction of neutrino-induced muons and the influences of the two media on angular and energy resolution are discussed in Ref. \cite{chiarusi}.

The ANTARES detector \cite{antares} is the largest underwater neutrino telescope. It consists of a three dimensional array of photomultiplier tubes located in the Mediterranean Sea. 
Initial results on the search for high energy cosmic neutrinos can be found elsewhere \cite{antares_diffuse,ANTAps2012}.

This paper reports on the measurement of the atmospheric $\nu_\mu$ energy spectrum in the range 100 GeV-200 TeV with the ANTARES neutrino telescope. Data collected from 2008 and 2011 have been analysed using a blinded procedure with two different analyses.

\section{Atmospheric neutrinos} \label{atmonu}
Cosmic rays are high energy particles, mostly protons and nuclei, arriving at the Earth. Their energy spectrum follows a power law, $\propto E^{-\gamma_p}$, where the spectral index $\gamma_p\simeq 2.7$ up to $\sim 10^{15}$ eV.
When cosmic rays enter the Earth's atmosphere they collide with atmospheric nuclei (mainly nitrogen and oxygen) and produce cascades of secondary particles.

Up to $\sim 100$ TeV, muons and neutrinos are produced mainly by decays of charged pions and kaons in the cascade and their spectra are related by the kinematics of the $\pi\rightarrow \mu \nu$ and $K\rightarrow \mu\nu$ decays. Additional lower energy neutrinos are produced by the muon decays.
The corresponding $\nu_\mu$ flux is usually referred to as the \textit{conventional atmospheric neutrino flux} and its intensity is expressed as
\begin{equation}\label{eq:gaisser} 
\frac{d\Phi_\nu} {dE_\nu d\Omega} (E_\nu,\theta) =  A_\nu E_\nu^{-\gamma_p} \biggl( \frac{ 1 }{ 1+\frac{aE_\nu}{\epsilon_\pi } \cos\theta} + \frac{ B }{ 1+\frac{bE_\nu}{\epsilon_K } \cos\theta } \biggr) \ ,
\end{equation}
in units of cm$^{-2}$s$^{-1}$sr$^{-1}$GeV$^{-1}$.
The \textit{scale factor} $A_\nu$, the \textit{balance factor} $B$ (which depends on the ratio of muons produced by kaons and pions) and the $a, b$ coefficients are parameters which can be derived from Monte Carlo computation, numerical approximations or from experimental data. The quantity $\epsilon_i$ (the \textit{characteristic decay constant}) corresponds to the energy at which the hadron interaction and decay lengths are equal. For pions and kaons, $\epsilon_\pi=115$ GeV and $\epsilon_K=850$ GeV respectively. 
An analytic description of the neutrino spectrum above 100 GeV is given by Volkova \cite{volkova}. Conventional atmospheric neutrino fluxes are also provided by the Bartol \cite{Gaisser,barr} and Honda \cite{honda} calculations. 
The expected power-law spectrum of conventional atmospheric neutrinos for $E_\nu \gg \epsilon_\pi,\epsilon_K$ can be approximated with 
\begin{equation}\label{eq:gaisser2} 
\frac{d\Phi_\nu} {dE_\nu } (E_\nu) =  A^\prime_\nu E_\nu^{-\gamma_\nu} \ ,
\end{equation}
where $\gamma_\nu\simeq \gamma_p +1$.

The major uncertainties in the calculations of the atmospheric neutrino flux arise from uncertainties on the composition, absolute normalisation and slope $\gamma_p$ of the primary cosmic ray spectrum, as well as the treatment of hadronic interactions in the particle cascades in the atmosphere. The uncertainty on the normalisation of the conventional atmospheric neutrino flux is estimated to be at the level of 25-30\% \cite{honda,bartolUnc}.

Charmed hadrons, produced by interactions of primary cosmic rays with air nuclei, have a much shorter  lifetime, approximately 5 to 6 orders of magnitude smaller than pions and kaons. This allows them to decay instead of interact, therefore producing a harder neutrino energy spectrum (\textit{prompt neutrino flux}). There is a significant variability in the different calculations of the prompt neutrino flux \cite{costa,martin,enberg} depending on the modelling of the hadronic interactions, the choices of gluon distributions and the renormalisation and factorisation scales.

\section{The ANTARES detector and the events reconstruction} \label{ant}

The ANTARES detector \cite{antares} is located at a depth of 2475 m in the Mediterranean Sea, 40 km offshore from Toulon, France (42$^\circ$48$^\prime$ N, 6$^\circ$10$^\prime$ E).
The full detector was completed in May 2008 and has been operating continuously ever since.
The telescope consists of 12 detection lines with 25 storeys each. A standard storey includes three optical modules (OMs) \cite{antares_om} each housing a 10-inch photomultiplier tube (PMT) \cite{antares_pmt} and a local control module that contains the electronics \cite{antares_electronics,antares_daq}. The OMs are orientated 45$^\circ$ downwards in order to optimise their acceptance to upgoing light and to reduce the effect of sedimentation and biofouling. The length of a line is 450 m and the horizontal distance between neighbouring lines is 60-75 m. The total number of active OMs is 885. 
The lines are connected to a junction box, which is connected to a shore station with a 42 km-long electro-optical cable. Through this cable the detector is powered, the data are collected and a clock signal, responsible for the synchronisation of the different detector elements, is distributed.

An accurate position calibration is required due to line displacement by the sea current. The shape of the lines and the orientations of the storeys are determined by an acoustic calibration system with tiltmeters and compasses placed on various storeys of the detector. The position of each optical module is determined with an accuracy better than 10 cm \cite{antares_pos}. The time offsets of the individual OMs were determined in dedicated calibration facilities onshore and are regularly monitored \textit{in situ} by means of optical beacons distributed in the apparatus \cite{antares_beacon}. A sub-nanosecond accuracy on the relative timing is achieved \cite{antares_time}. 

In the ANTARES convention, upgoing events have zenith angle $\theta>90^\circ \ (\cos\theta<0)$, while downgoing events (dominated by atmospheric muons) have $\theta\le 90^\circ\ (\cos\theta \ge 0)$.
A high-energy $\nu_\mu$ that interacts in the matter around or within the detector produces a relativistic muon that can travel hundreds of metres and cross the detector or pass nearby. 
This muon induces Cherenkov light when travelling through the water, which is detected by the OMs. From the time and position information (hit) of the photons recorded by the OMs, the energy and direction of the muon is reconstructed. These quantities are correlated to the parent neutrino energy and direction.
Since atmospheric muons cannot traverse the Earth, a directional cut selecting upgoing tracks significantly reduces this background. 

The algorithm used to reconstruct the muon direction uses four consecutive steps for the fitting procedure. 
The first three steps -- a linear $\chi^2$-fit, an ``M-estimator'' minimisation and a simplified likelihood fit -- provide a starting point for the last likelihood fit. 
The signal hit selection is purely based on coincidences and on causality criteria. These criteria require that the distance between different OMs must be related to the distance travelled by the light in the medium within the observed time difference. The final likelihood is based on a probability density function of the hit time residuals, defined as the time differences between the observed and expected hits on the optical modules. Hits due to optical background and Cherenkov light from secondary particles, as well as light scattering are taken into account. The variable $\Lambda$, defined as
\begin{equation}  \label{Lambda_def}
  \Lambda \equiv \frac{\log L}{N_{hit}-5} + 0.1 (N_{comp} -1) \ ,
\end{equation}
is used to characterize the quality of the fit. The first term is the log-likelihood value per degree of freedom of the fit, $i.e.$ the number of hits, $N_{hit}$, used in the fit minus the number of fit parameters. These five parameters are the local zenith and azimuth angles, and the impact coordinates of the track on an ideal cylinder surrounding the detector's instrumented volume.
$N_{comp}$ is the number of starting points of the ``M-estimator'' that result in a track direction within 1$^{\circ}$ from the result with the best likelihood per degree of freedom. In most cases, $N_{comp}=1$ for badly reconstructed events, while it can be as large as nine for well reconstructed events. The coefficient 0.1 in q. (\ref{Lambda_def}) was chosen via Monte Carlo simulations to maximise the separation in $\Lambda$ between simulated signal and misreconstructed downgoing muons.
The algorithm does not use any hit amplitude information. 

The reconstruction quality parameter $\Lambda$ is negative and takes values closer to zero for well reconstructed tracks. This parameter can be used to reject atmospheric muons that have been misreconstructed as upgoing. 
In addition, the fit algorithm provide an estimate of the angular uncertainty on the muon track direction, $\beta$, which is used as an additional quality parameter to further reject misrecontructed atmospheric muons.
There is good agreement between simulation and data for the cumulative distribution of the reconstruction quality variable $\Lambda$ for upgoing tracks which have an angular error estimate $\beta<1^\circ$ as reported in Ref. \cite{ANTAps2012}.
The median angular resolution of selected events from simulated cosmic neutrinos is $0.46^\circ\pm 0.10^\circ $ and 83\% are reconstructed within $1^\circ$ of the true neutrino direction. 

The measurement of the neutrino energy is a non trivial problem. The events considered in this analysis are almost completely passing-through muons, generated outside the detector and traversing it. Only a fraction of the neutrino energy is transferred to the detected muon, which is often produced outside the instrumented volume of the detector. In addition, as the muon travels, it loses energy before being detected. 

Muon energy losses are usually classified into continuous and discrete processes. The former is due to excitation/ionisation, which depends weakly on muon energy and can be considered nearly constant for relativistic particles. For muons below $\sim 500$\,GeV, this is the dominant energy loss process. At higher energies, discrete energy losses become important:~bremsstrahlung, direct electron-positron pair production and electromagnetic interaction with nuclei. In these processes energy is lost in bursts along the muon path. In general, the total muon energy loss is parameterized as 
\begin{equation}\label{eq:muloss} 
\frac{\textrm{d}E_\mu}{\textrm{d}X} = -\alpha-\beta E_\mu \ ,
\end{equation}
where $X$ is the thickness of crossed material, $\alpha$ accounts for the excitation/ionisation energy loss and $\beta$ for the three mentioned radiation energy loss processes. 
The coefficients $\alpha$ and $\beta$ in Eq.~(\ref{eq:muloss}) are mildly energy dependent as well as dependent upon the chemical composition of the medium:~in particular $\alpha\propto Z/A$ and $\beta \propto Z^2/A$. Typical values of the $\alpha(E)$, $\beta(E)$ coefficients in water are reported in Ref. \cite{Mu_Eloss}. 

Along with Cherenkov light emission, muons travelling in water produce hadronic and electromagnetic showers because of radiative energy loss processes, and additional light is produced by the secondary particles. The amount of detected light can be used to infer the energy of the muon. This information can be subsequently used to determine the energy of the parent neutrino. The neutrino energy distribution is distorted by the limited energy resolution and the overall acceptance of the detector. The measured muon energy distribution is translated into the atmospheric neutrino energy spectrum through a response matrix, determined from Monte Carlo simulations, and an unfolding procedure.

\section{Neutrino energy estimation} \label{energyest}

The methods used to reconstruct the muon energy are based on the amount of detected light on the OMs.
The muon estimated energy was determined for each event; the parent neutrino distribution was derived with unfolding procedures, as discussed in \S \ref{unfolding}.
The expected number of photoelectrons on each OM, $\langle n_{pe}\rangle$, is a function of the muon energy, water properties, of the detector configuration and OM distance and  orientation from the light source. $\langle n_{pe}\rangle$ is calculated considering the amount of light emitted while a muon traverses the detector, taking into account contributions from \textit{direct} and \textit{scattered} light. Direct photons are those originating along the muon trajectory and arriving on OMs in the Cherenkov wavefront without being scattered. 
Scattered photons are delayed by the increased optical path from the emission point to the OM. 
Above $\sim 500$ GeV most of the Cherenkov light emitted along the muon path comes from the secondary particles produced in radiation losses. 
The total amount of light emitted from the muon and collected by the OMs is directly correlated to its energy.

Two independent methods are used in this work to estimate the muon energy. The first one -denoted in the following as \textit{energy likelihood method} \cite{dimitri} - maximises the agreement of the expected amount of light in the optical modules with the amount of light that is actually observed. Starting from the direction information of the track reconstruction procedure and keeping the energy of the muon $E_\mu$ as a free parameter, a maximum likelihood function is constructed as
\begin{equation}
  \mathcal{L}(E_\mu) = \frac{1}{N_{OM}} \prod_{i}^{N_{OM}}\mathcal{L}_i(E_\mu) \ .
  \label{Likelihood_def}
\end{equation}
This product is taken over all the $N_{OM}$ optical modules positioned up to 300 m from the reconstructed track, regardless of whether a hit was recorded or not. Optical modules with unusually high or low counting rates in a particular run, as well as those that are not operational, are excluded. $\mathcal{L}_i(E_\mu)$ depends on the probability of observing a pulse of measured amplitude $Q_i$ given a certain number of photoelectrons produced on the $i^{th}$ OM. These individual likelihood functions $\mathcal{L}_i(E_\mu)$ are constructed as
\begin{subequations}
\begin{equation}
  \mathcal{L}_i(E_\mu) \equiv P(Q_i; \langle n_{pe} \rangle) = \sum_{n_{pe}=1}^{n_{pe}^{max}} P(n_{pe}; \langle n_{pe} \rangle) \cdot G(Q_i; n_{pe}) \ ,
  \label{Likelihood_ind}
\end{equation}
when a hit is recorded and
\begin{equation}
  \mathcal{L}_i(E_\mu) \equiv P(0; \langle n_{pe} \rangle) = e^{-\langle n_{pe}\rangle}+P_{th}(\langle n_{pe}\rangle) \ ,
  \label{Likelihood_ind_0}
\end{equation}
\end{subequations}
when there is no hit on the optical module. Equation (\ref{Likelihood_ind}) consists of two terms, the Poisson probability $P(n_{pe}; \langle n_{pe} \rangle)$ of having $n_{pe}$ photoelectrons given an expectation of $\langle n_{pe} \rangle$, and a Gaussian term $G(Q_i; n_{pe})$ which expresses the probability that $n_{pe}$ photoelectrons on the photocathode will yield the measured amplitude $Q_{i}$. Equation (\ref{Likelihood_ind_0}) consists of a term describing the Poisson probability of observing zero photoelectrons when the expected value is $\langle n_{pe} \rangle$, and a term, $P_{th}(\langle n_{pe}\rangle)$, describing the probability that a photon conversion in the optical module will give an amplitude below the threshold level of 0.3 photoelectrons.

The second muon energy estimation method - denoted in the following as \textit{energy loss method} \cite{fabian} - relies on the muon energy losses along its trajectory, Eq. (\ref{eq:muloss}). The muon energy deposit per unit path length is approximated by an estimator $\rho$ which can be derived from measurable quantities
\begin{equation}  \label{dEdX_def}
  \frac{dE}{dX} \propto \rho = \frac{\sum_{i=1}^{N_{hit}}Q_i}{\epsilon}\cdot\frac{1}{L_\mu} .
\end{equation}
The quantity $L_\mu$ represents the length of the reconstructed muon path starting from the entry point on the surface of the cylinder surrounding the instrumented volume of the detector. Due to the light transmission properties of the water, this volume is defined extending the radius and height of the cylinder by twice the light attenuation length. 
$L_\mu$ is thus longer than the effective visible track in the detector. 
$Q_i$ is, as before, the measured amplitude of the $i$-th OM. To remove the contribution from background light using the causality criteria embedded in the reconstruction algorithm, only the hits used in the final tracking step are considered. Finally, the quantity $\epsilon$ represents the overall ANTARES light detection capability. This quantity depends on the geometrical position and direction of the muon track. It is derived on an event-by-event basis as
\begin{equation}
  \epsilon = \sum_{i=1}^{N_{OM}}\exp\left(-\frac{r_i}{\lambda_{abs}}\right) \cdot \frac{\eta_i(\vartheta_i)}{r_i}.
  \label{dEdX_epsilon}
\end{equation}
Here the sum runs over all the active optical modules. The distance from the muon track, $r_i$, and the photon angle of incidence, $\vartheta_i$, are calculated for each OM; $\vartheta_i$ is used to obtain the corresponding angular acceptance $\eta_i(\vartheta_i)$ \cite{antares_om} of the involved OM. The distance $r_i$ is used to correct for the light absorption in water (with characteristic absorption length $\lambda_{abs}=55$ m) taking into account the light distribution within the Cherenkov cone.

\section{Energy spectrum unfolding} \label{unfolding}

Due to the steeply falling neutrino energy spectrum and the uncertainty in the event reconstructed energy, an unfolding procedure has to be used in order to draw the actual energy spectrum from the distribution of the measured event-by-event measured. This procedure has to take into account the stochastic nature of the muon energy losses, the large uncertainty in the reconstructed energy, the detection inefficiencies and the fact that only the daughter muon energy is measured.

The problem to be solved is a set of linear equations of the form
\begin{equation}  \label{Axa}
A\mathbf{e}=\mathbf{x} \ .
\end{equation}
Vector $\mathbf{e}$ represents the true unknown distribution in a discrete number of intervals, vector $\mathbf{x}$ is the measured distribution and the matrix $A$, called the response matrix, is the transformation matrix between these two distributions. The response matrix is built using Monte Carlo simulations.

A simple direct inversion of the response matrix leads in most cases to a rapidly oscillating solution and large uncertainties due to the fact that the matrix $A$ is ill-conditioned \cite{regularisation}: minor fluctuations in the data vector $\mathbf{x}$ can produce large fluctuations in the solution $\mathbf{e}$.

One of the methods used to solve this problem is the \textit{singular value decomposition} \cite{SVDUnfolding}. The response matrix is decomposed as $A=USV^T$, where $S$ is a diagonal matrix and $U$ and $V$ are orthogonal matrices. This is equivalent to expressing the solution vector $\mathbf{e}$ as a sum of terms weighted with the inverse of the singular values of the matrix $S$. However, small singular values can enhance the statistically insignificant coefficients in the solution expansion, leading to the same problem appearing when directly inverting the response matrix. A way to overcome this problem is to impose an external constraint on how the solution is expected to behave. The process of imposing such a constraint is called regularisation. The constraint used in our procedure (\S \ref{analysisA}) and described in Ref. \cite{SVDUnfolding} controls the curvature of the solution, not allowing vector $\mathbf{e}$ to exhibit large bin-to-bin fluctuations, which are not physically motivated.

Another unfolding method which does not rely on the regularisation procedure, is the iterative method based on \textit{Bayes' theorem} in Ref. \cite{BayesUnfolding}. 
The Bayes' theorem states that the probability $\mathcal{P}(E_i|X_j)$ that $E_i$ is the true energy, given the measurement of a value $X_j$ for the energy estimator, is equal to
\begin{equation}
  \mathcal{P}(E_i|X_j) = \frac{A(X_j|E_i)p_0(E_i)}{\sum_{l=1}^{n_E}A(X_j|E_l)p_0(E_l)} \ ,
\label{BayesTh}
\end{equation}
where $A(X_j|E_i)$ is the probability (calculated from Monte Carlo simulations) of measuring an estimator value equal to $X_j$ when the true energy is $E_i$. This quantity corresponds to the element $A_{ij}$ of the response matrix. The \textit{a priori} probability $p_0(E_j)$ is the expected energy distribution at the detector derived from theoretical expectations and Monte Carlo simulations. For a given estimator distribution, the energy distribution at the detector can be obtained applying Eq. \ref{BayesTh} iteratively. At the $n$-th iteration, the energy distribution at the detector $p_n(E_j)$ is calculated taking into account the observed number of events in the estimator distribution and the expectations from $p_{n-1}(E_j)$. The result rapidly converges towards a stable solution. The number of iterations to be performed is optimised by applying the procedure to different Monte Carlo samples and studying the convergence of the obtained solutions. A small number of iterations biases the unfolding result towards the prior probability $p_0(E_j)$, while further iterations  beyond the point where convergence is reached enhances statistical fluctuations in the solution.

Both mentioned unfolding procedures are used in this analysis and are implemented in the \textit{RooUnfold} package \cite{roounfold} as part of the ROOT framework \cite{ROOT}.

\section{Data analysis} \label{procedure}

\subsection{The data sample and Monte Carlo simulations } \label{data}

The analysed sample covers ANTARES data acquired from December 2007 to December 2011. It comprises an equivalent livetime of 855 days. 
For each data run a similar Monte Carlo run of atmospheric muons and neutrinos is generated. In the simulated files, the specific conditions of the detector and of the environment are reproduced.
The optical background in the OMs is added by sampling the count rate from the real data, in order to ensure that the simulation contains the same background as the analysed data. 

The simulation starts with the generation of upgoing neutrino events, using the GENHEN package \cite{genhen} which uses CTEQ6D \cite{cteq6d} parton density functions to compute the neutrino cross-section. Events are weighted according to the cross-section and their probability to traverse the Earth. If the neutrino interaction occurs near the detector, the resulting hadronic shower is simulated. If the interaction is external to the detector volume, only the resulting muon is propagated towards the detector. Downgoing atmospheric muons, the main source of background for the analysis, are simulated with the MUPAGE program \cite{muparam,mupage}, which provides a parameterised flux of muons in bundles at the detector. 
Cherenkov photons are simulated inside the detector by sampling tabulated values of photon arrival times, taking into account the measured absorption and scattering parameters. 

The two unfolding methods described in \S \ref{unfolding} are applied on data reconstructed using the energy estimators described in \S \ref{energyest}. 
Both methods require high purity to avoid corrupting the final result by the presence of wrongly-reconstructed atmospheric muons mimicking upgoing neutrino-induced events. 
All cuts are optimised on Monte Carlo simulations. A 10\% fraction of the data is initially used to check the agreement between the observed and expected quantities both on downgoing (atmospheric muons) and upgoing (atmospheric neutrinos) events.
The remaining 90\% of the data set is \textit{unblinded} only when all the cuts and optimisation procedures have been defined.

\subsection{Unfolding method based on Bayes' theorem} \label{analysisB}
The energy of the muon reconstructed using the energy {energy loss method} is used to derive the neutrino energy spectrum through the unfolding method based on Bayes' theorem.
In order to suppress wrongly-reconstructed atmospheric muons, the reconstruction quality parameter $\Lambda$ and the angular error estimate $\beta$ are fixed to $  \Lambda > -4.9\; , \;\beta < 1.0^{\circ}$ and the angular region is restricted to $\theta > 100^{\circ}$. 
The response matrix is built by weighting Monte Carlo events according to the flux from Ref. \cite{barr} with no prompt contribution.
The expected atmospheric neutrino rate is 1.8 events per day with a contamination from wrongly reconstructed atmospheric muons of $\sim0.3\%$ \cite{luigi}. 
\begin{figure}
  \centering
\vskip -2cm
  \includegraphics[width=0.6\textwidth]{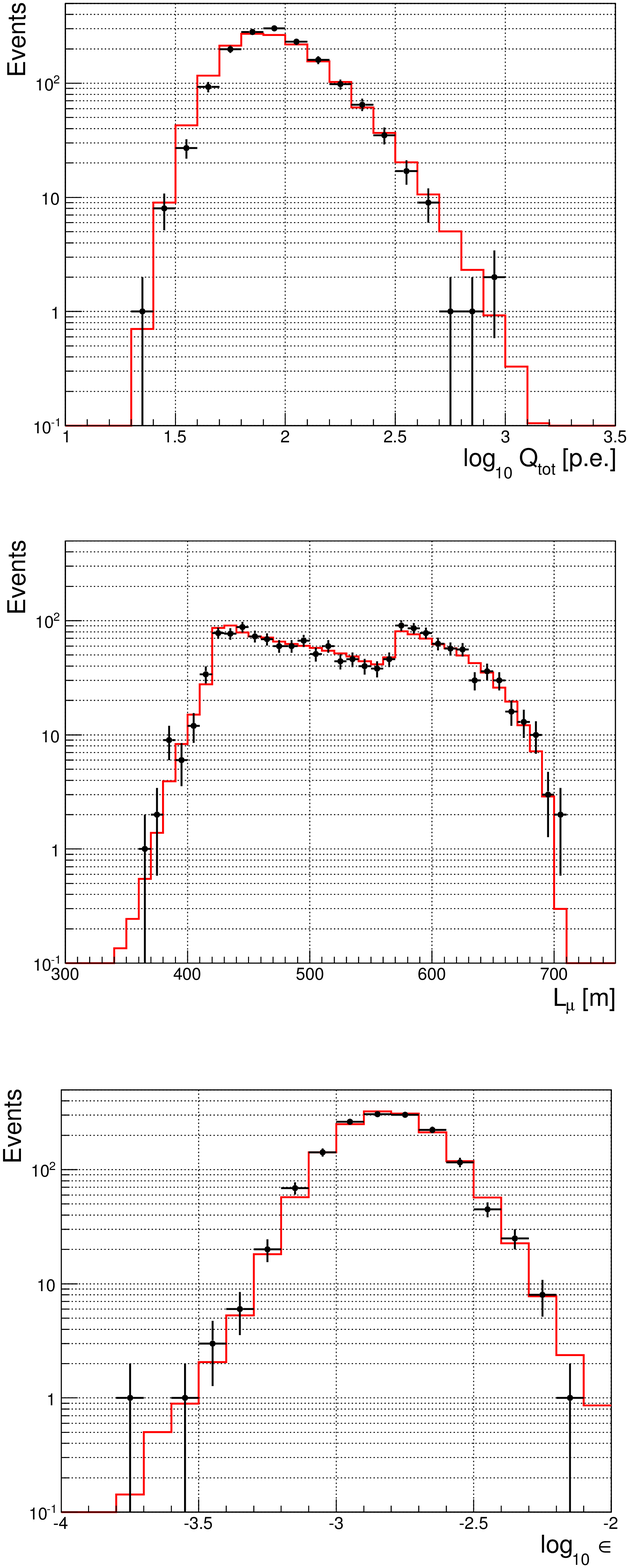}
  \caption{Comparison between data (black crosses) and simulations (red line) for the quantities used to construct the energy loss estimator $\rho$, Eq. \ref{dEdX_def}. (Top) Distribution of the total measured amplitude $Q_{tot}$ (in photoelectrons) on the optical modules involved in the events; (Middle) muon track length in the detector region; (Bottom) light collection capability defined in Eq. \ref{dEdX_epsilon}.
Only events passing the selection criteria of the \textit{energy loss method} are drawn. The Monte Carlo prediction is scaled by a factor 1.25.}
  \label{fig:dataMC_dEdX}
\end{figure}

The precision on the reconstructed energy $ E_\mu^{Rec}$ depends on the true muon energy $E_\mu^{MC}$. The quantity
\begin{equation}\label{ereso}
\delta E_{\mu} \equiv \log_{10} \frac{E_\mu^{Rec}}{E_\mu^{MC}} 
\end{equation}
is used to estimate the energy resolution of the reconstruction. The standard deviation of a Gaussian fit for different intervals of the Monte Carlo true muon energy achieved with this method is almost constant at $\sigma_{\delta E_{\mu}}\simeq 0.4$ over the considered energy range.

The comparison between the distribution of the quantities used to construct the energy estimator $\rho$ (Eq. \ref{dEdX_def}) for data and Monte Carlo events is shown in Fig. \ref{fig:dataMC_dEdX}. The overall predicted number of Monte Carlo events is $\sim$25\%  lower than the measurement, within the expected flux normalisation uncertainty. 

The relation between the distribution of the neutrino energy and the measured observable $\rho$ is described by the response matrix constructed via Monte Carlo. The iterative unfolding method based on Bayes' theorem moves the distribution of the observed estimator towards the real neutrino energy distribution starting from an \textit{a priori} hypothesis. 
The optimal value of the number of iterations is established using a $\chi^2$ test on different pseudo-data sets, which are unfolded for different number of iterations. The atmospheric neutrino flux from Ref. \cite{barr} is used as the \textit{a priori} spectrum to construct the response matrix. The spectral index corresponds to the parameter $\gamma_\nu$ in Eq. \ref{eq:gaisser2} and it assumes the value $\gamma_\nu=3.63$ when the neutrino flux is averaged over the lower hemisphere.
The optimal number of iterations is found to be equal to five using a $\chi^2$ test comparing the unfolded result and the true neutrino spectrum of pseudo-data samples. In particular, neither an enhancement of statistical fluctuations deriving from a larger number of iterations, nor a bias towards the \textit{a priori} spectrum used to construct the response matrix is observed.

\begin{figure}
\centering
\includegraphics[width = 0.8\textwidth]{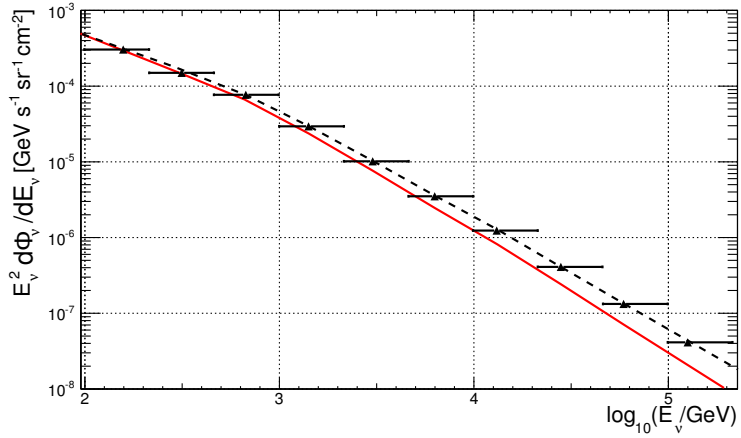}
\caption{Unfolding of a known spectrum. The red solid line is the energy spectrum from Ref. \cite{barr} used as the \textit{a priori} probability for the bayesian unfolding of pseudo-data generated according to an injected spectrum (black dashed line). The unfolding result (black symbols) is shown without error bars.}
\label{fig:unfolding_iterations}
\end{figure}

Figure \ref{fig:unfolding_iterations} shows the result of the unfolding procedure on a pseudo-data set with energy spectrum flatter by a factor $E_\nu^{+0.1}$ with respect to the \textit{a priori} spectrum. 
This pseudo-data set has an overall normalisation 20\% larger than the \textit{a priori} one, more in agreement with the measured number of events in the data.
The points in Fig. \ref{fig:unfolding_iterations} represent the result of the unfolding of this pseudo-data set. The deviations between the true distribution and the unfolded one will be considered in the discussion of systematic uncertainties on \S \ref{uncertainties}

\subsection{Unfolding method based on singular value decomposition} \label{analysisA}

The muon energy reconstructed using the {energy likelihood method} is used to build the vector $\mathbf{x}$ in Eq. (\ref{Axa}) of the singular value decomposition unfolding method.
Here, the cut on the reconstruction quality parameter $\Lambda$ is the same as in the {energy loss method}  ($\Lambda > -4.9$). The cut on $\beta$ is slightly more stringent ($\beta < 0.5^{\circ}$), but the zenith angle region is larger, as only events with $\theta < 90^{\circ}$ are rejected.
The response matrix is built weighting Monte Carlo events according to the conventional flux from Ref. \cite{barr} plus the prompt contribution from Ref. \cite{enberg}.
The corresponding neutrino event rate is 1.7 events per day and the expected muon contamination below 0.2\%.
The estimated energy resolution Eq. (\ref{ereso}) achieved with this method improves from $\sigma_{\delta E_{\mu}}\simeq 0.45$ at $E_\mu^{MC}=500$ GeV to $\sigma_{\delta E_{\mu}}\simeq 0.3$
when $E_\mu^{MC}=10^3$ TeV.

\begin{figure}
  \centering
  \includegraphics[width=0.8\textwidth]{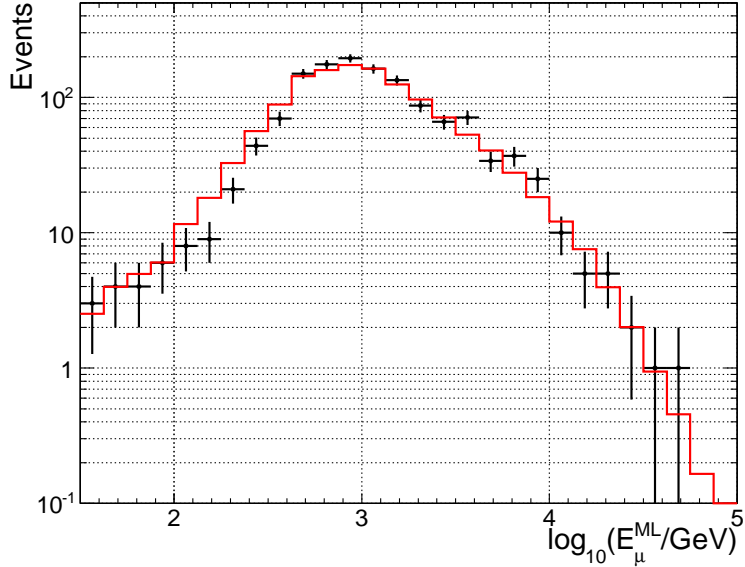}
  \caption{Data (black crosses) and Monte Carlo (red line) comparison for the maximum likelihood muon energy estimator $E^{ML}_{\mu}$, for events passing the selection cuts. This distribution is obtained at the end of the procedure, after the data unblinding (see text). The Monte Carlo prediction is scaled by a factor 1.25. }
  \label{fig:dataMC_ML}
\end{figure}

Figure \ref{fig:dataMC_ML} shows the comparison between data and simulation of the muon reconstructed energy using the  {energy likelihood method} for events passing the selection (simulation events are normalised to the data). Also in this case, the simulation prediction is $\sim$25\% lower than data. 

The singular value decomposition unfolding procedure necessary to derive the neutrino energy spectrum at the detector is applied to this distribution.
The result of the unfolding is dependent on the choice of the regularisation parameter, i.e. how strong the regularisation condition acts in smoothing unexpected oscillating components due to statistical fluctuations. 
A large value of the regularisation parameter imposes stronger constraints on the solution with a possible bias towards the assumed underlying spectrum. 
The regularisation parameter is chosen by examining the distribution of the absolute values of the expansion coefficients, as described in Ref. \cite{SVDUnfolding}. 
The values of the expansion coefficients drop rapidly as the singular values decrease, reaching a level where they are compatible with zero, i.e. following a normal distribution with zero mean and standard deviation equal to one. 
The optimal regularisation parameter is equal to the square of the singular value that corresponds to the coefficient above which the remaining values are compatible with zero. This behaves as a Fourier low pass filter, progressively damping out insignificant terms in the solution expansion. Similarly to the Bayesian unfolding method, the performance of the singular value decomposition unfolding was tested on Monte Carlo samples, with similar results as those shown in Fig. \ref{fig:unfolding_iterations}.

\section{$\nu_\mu$ energy spectrum measurement } \label{merging}
\begin{figure}
  \centering
  \includegraphics[width=\textwidth]{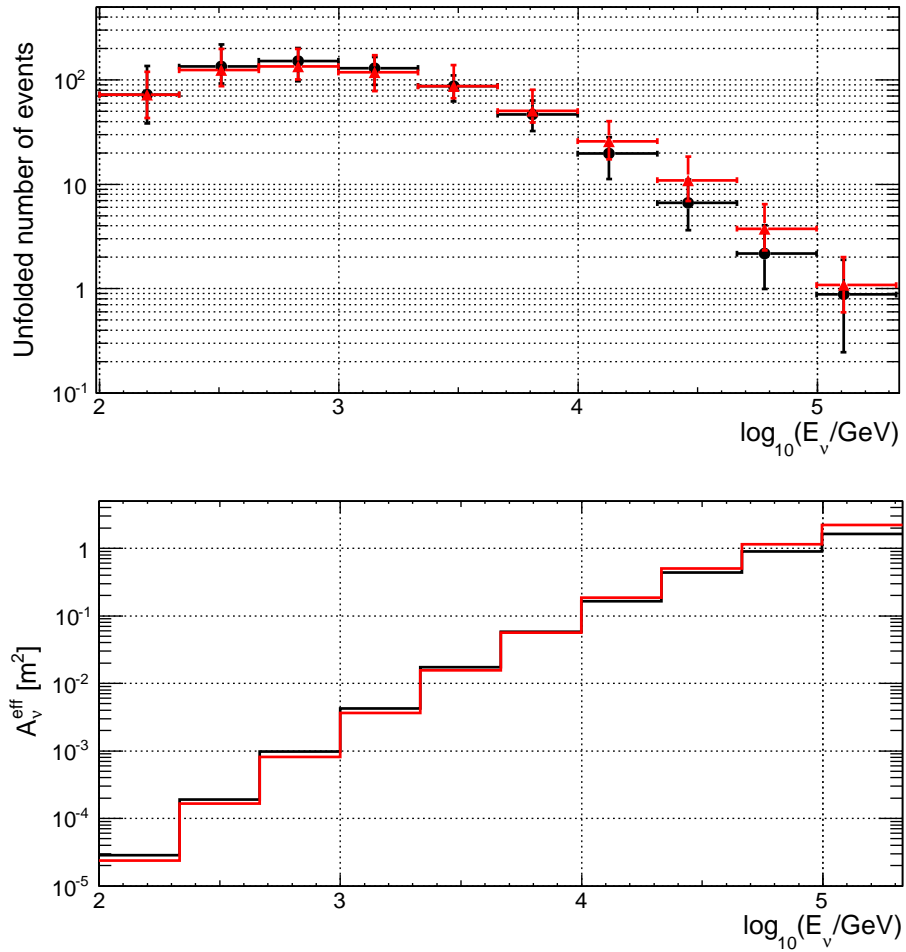}
  \caption{Top: unfolded energy distribution at the detector for the  {energy likelihood} (red) and the  {energy loss} (black) methods. The numbers correspond to events per bin per year of effective livetime. Bottom: corresponding neutrino effective area for upgoing neutrinos for the two methods.}
  \label{fig:aeff}
\end{figure}

The output of the unfolding represents a detector-dependent quantity, as it corresponds to the number of events per energy bin in the considered livetime. The top panel of Fig. \ref{fig:aeff} shows the neutrino energy distribution at the detector resulting from the two methods. These energy distributions are dependent on the selection criteria and on the analysed solid angle which are different for the \textit{energy likelihood} and \textit{energy loss} methods.

A detector-independent spectrum is derived taking into account the detection and selection efficiencies of the apparatus as a function of the incoming neutrino zenith angle and energy. These effects are included in the so-called neutrino effective area of the detector, $A^\textrm{eff}_\nu(\theta,E_\nu)$. As $A^\textrm{eff}_\nu$ depends on the neutrino cross-section, this quantity is different for $\nu_\mu$ and $\overline \nu_\mu$.
Here, $A^\textrm{eff}_\nu$ is defined as the ratio between the selected events and the atmospheric neutrino plus antineutrino flux as a function of the zenith angle $\theta$ and neutrino energy $E_\nu$. 
In addition to the neutrino cross-section, the neutrino effective area depends on the absorption of neutrinos through the Earth and on the muon detection and selection efficiency.
The bottom panel of Fig. \ref{fig:aeff} shows the neutrino effective area as a function of $E_\nu$ for the two methods used in this analysis. The differences between the two include the effects of the different quality cut on the reconstructed tracks.

The effective area is used to relate the energy distribution at the detector to the energy distribution at the surface of the Earth. A correction factor for the effective area takes into account the fact that the \textit{energy loss} method considers only events with $\theta> 100^\circ$.
\begin{figure}
  \centering
  \includegraphics[width=1.0\textwidth]{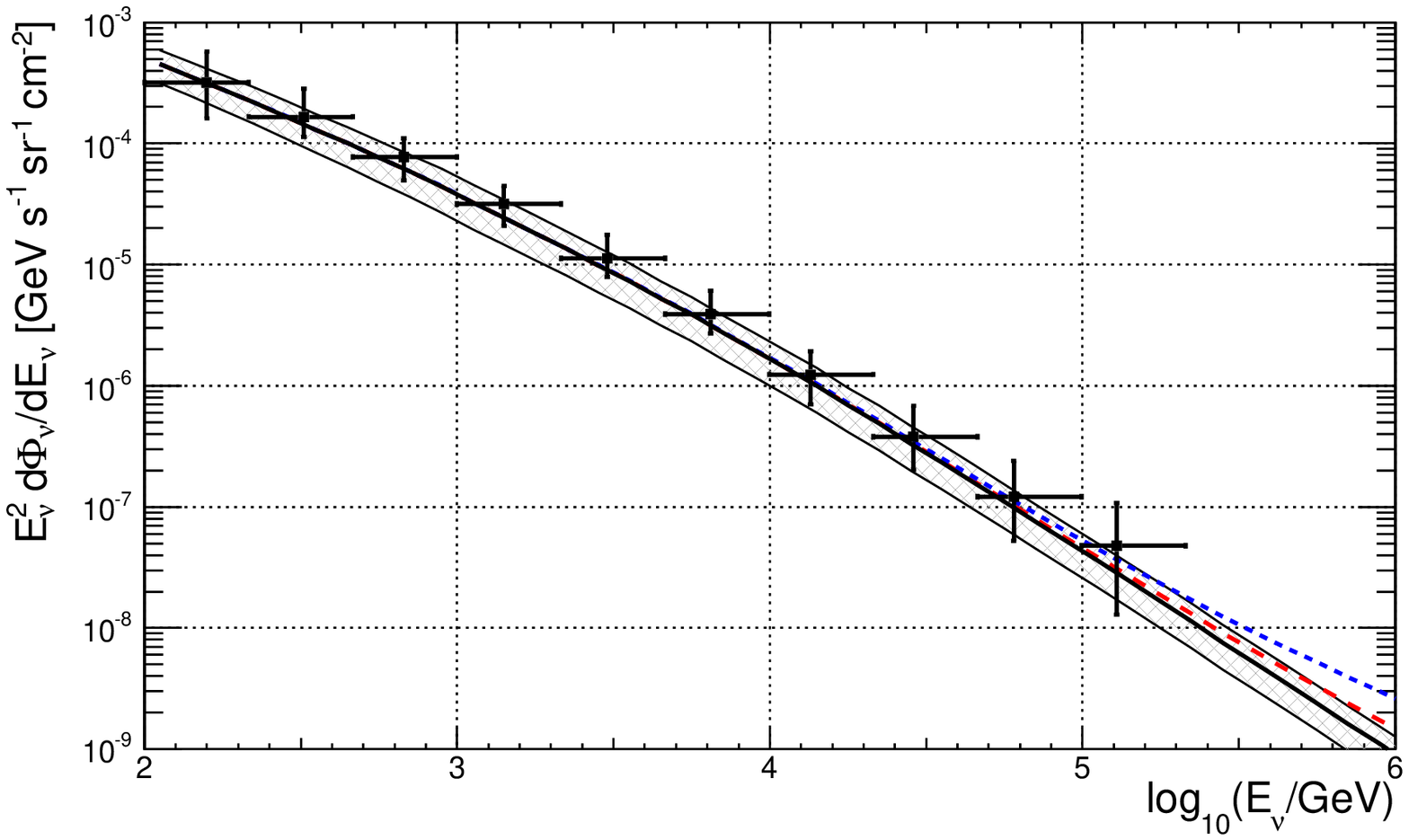}
  \caption{Atmospheric neutrino energy spectrum obtained with the ANTARES neutrino telescope using 2008-2011 data. The flux reported here is multiplied by E$^2$ and compared with the expectations from Ref. \cite{barr}. The gray band corresponds to the uncertainty in the flux calculation from Ref. \cite{bartolUnc}.
The flux obtained by adding to the conventional flux the prompt contributions from Ref. \cite{martin} (red - dashed line) and Ref. \cite{enberg} (blue - short dashed line) is also drawn.}
  \label{fig:SpectrumBartol}
\end{figure}

\begin{table} 
  \[
  \begin{array}{cccc}
    \hline
\text{Energy range}     & \log_{10}({\overline E}_{\nu}/\textrm{GeV})    &{\overline E_\nu}^2\cdot {d\Phi_\nu/dE_\nu}       &\%\text{ Uncertainty}  \\
\log_{10}({E}_{\nu}/\textrm{GeV})    &   &\text{[GeV s$^{-1}$ sr$^{-1}$ cm$^{-2}$]}&   \\
    \hline\hline
    2.00 - 2.33         &   2.20            &    3.2 \times 10^{-4}           &   -49, +80       \\
    2.33 - 2.66         &   2.51            &    1.7 \times 10^{-4}           &  -32 ,  +69 \\
    2.66 - 3.00         &   2.83            &    7.8 \times 10^{-5}           &  -36   ,    +41 \\ 
    3.00 - 3.33         &   3.15            &    3.2 \times 10^{-5}           &  -34  ,     +40 \\
    3.33 - 3.66         &   3.48            &    1.1 \times 10^{-5}           &  -30  ,     +55 \\
    3.66 - 4.00         &   3.81            &    3.9 \times 10^{-6}           &  -31  ,    +56 \\
    4.00 - 4.33         &   4.13            &    1.2 \times 10^{-6}           &  -43  ,    +56 \\
    4.33 - 4.66         &   4.46            &    3.8 \times 10^{-7}           &  -46  ,     +80 \\
    4.66 - 5.00         &   4.78            &    1.2 \times 10^{-7}           &  -57  ,    +96 \\
    5.00 - 5.33         &   5.11            &    4.8 \times 10^{-8}           &  -73  ,    +125 \\
    \hline
  \end{array}
  \]
  \caption{The unfolded atmospheric neutrino energy spectrum from the ANTARES neutrino telescope. Each row shows the energy range of the bin; the weighted central value of the neutrino energy $\overline E_\nu$ in the bin; the flux multiplied by $\overline E_\nu^2$ and the percentage uncertainty on the flux.}
  \label{tab:finalspectrum}
\end{table}

The weighted central value of the neutrino energy bin has been calculated taking into account the steep decrease of the energy spectrum. 
The flux is obtained by dividing the contents of the two histograms presented in Fig. \ref{fig:aeff} and averaging the results of the two methods. The measured atmospheric neutrino energy spectrum for $\theta>90^{\circ}$  is presented in Fig. \ref{fig:SpectrumBartol} and the values are reported in Tab. \ref{tab:finalspectrum}. 
The differences for each method with respect to their average value 
are much smaller than most of the considered systematic uncertainties (see next section) and are shown in Fig. \ref{fig:syst_errors} as a thin black line. 
The obtained flux values, with the estimated uncertainties, can be fitted according to Eq. \ref{eq:gaisser} in the analysed energy range. The resulting best fit value, corresponding to the neutrino spectral index for a power law behavior in the energy region where the assuptions of Eq. \ref{eq:gaisser2} are valid, is $\gamma_{meas}=3.58\pm 0.12$. This value is to be compared with $\gamma_{\nu}=3.63$ obtained when the \textit{a priori} pseudo-data set is used.

\section{Systematic uncertainties} \label{uncertainties}
The result of the unfolding process is dependent on Monte Carlo simulations via the construction of the response matrix. The simulations depend on a number of parameters with associated uncertainties that influence the unfolding result systematically. 
Most of the effects inducing systematic uncertainties on the measurement of the neutrino flux and energy have been already studied in Refs. \cite{antares_diffuse,ANTAps2012}. 

The impact of the variations of these parameters is estimated using different specialised neutrino simulation datasets, varying only one parameter each time. 
The simulation set obtained with the standard parameters, corresponding to our best estimate of those parameters, is used to construct the default response matrix. Each modified Monte Carlo sample was then used as pseudo-data and unfolded. The deviation in each energy bin from the spectrum obtained with the default value of the parameter corresponds to the systematic uncertainty associated with this parameter variation. 
The systematic uncertainties as a function of the neutrino energy are different for the two methods due to the different unfolding procedures and constructions of the energy estimators. 

\begin{figure}
\centering
\includegraphics[width = \textwidth]{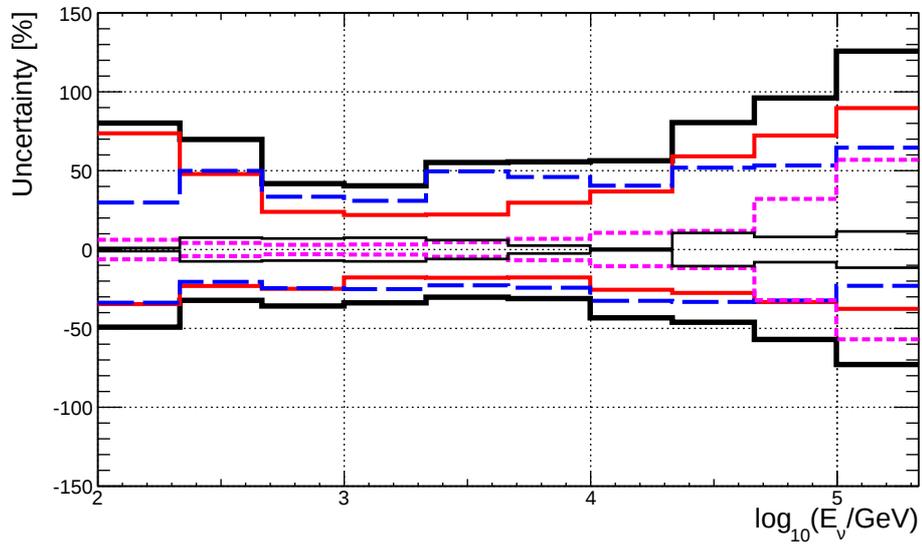}
\caption{Systematic uncertainties calculated for each neutrino energy bin. Red continuous line represents the effects given by a $\pm 10\%$ change of the OM efficiency with respect to the default value; blue dashed line is for a $\pm 10\%$ change in the absorption length in water;
magenta short dashed line is the statistical uncertainties. 
The thick black line shows the estimated total uncertainty while the thin black line represents the relative difference between the two unfolding results. The effects from the OM angular acceptance and the change in the underlying event weighting model (see text) are not shown in the figure.}
\label{fig:syst_errors}
\end{figure}

Figure \ref{fig:syst_errors} shows the percentage uncertainty as a function of the binned neutrino energy with respect to the corresponding value of the flux reported in Fig. \ref{fig:SpectrumBartol}.
The largest uncertainty arising from the \textit{energy loss} or from the \textit{energy likelihood}  methods is considered in each bin.
The total uncertainty (black full line) is computed as the quadratic sum of each contribution, separately for positive and negative deviations. 
The differences between the spectra obtained with the two energy estimators with respect to the average value is shown as the thin black line. 

The overall sensitivity of the optical modules (red continuous line) has been modified by $\pm10\%$. This includes the uncertainty on the conversion of a photon into a photoelectron on the PMT photocathode as well as other effects related to the OM efficiency.
The value of $\pm 10$\% was obtained from a study of the $^{40}$K decay rate observed in the detector \cite{ANTAk40} and the rate and zenith angle distribution of detected atmospheric muons \cite{ANTA5lines}.

A second uncertainty related to the optical modules is that connected with the angular acceptance, i.e. the angular dependence of the light collecting efficiency of each OM. Two different response curves, centred on the nominal one and departing from it in opposite directions, have been used as input of the dedicated Monte Carlo simulation. This affects the measurement by less than 10\% over the whole analysed energy range. 

The uncertainties on water properties were studied in Ref. \cite{ANTA5lines} and are taken into account by scaling up and down by 10\% the absorption length of light in water with respect to the nominal value (blue dashed line).

The effects due to the uncertainty in the neutrino flux used in the response matrix of the unfolding procedures include the possible contribution of prompt neutrinos \cite{martin}, the effect of a slope change of $\pm 0.1$ in the \textit{a priori} spectral index $\gamma_\nu$ and the effect due to the chosen number of iterations (see \S \ref{analysisB}). The uncertainties deriving from these effects, not shown in Fig \ref{fig:syst_errors}, are always below 10\%.

The unfolding result is influenced by the energy estimators, as well as by the unfolding method and the event selection criteria. In particular the \textit{energy likelihood} method is more sensitive to variations of water properties, while the \textit{energy loss} method has a larger dependence on OM efficiency. 
The statistical uncertainty (magenta short dashed line) is relevant only for the highest energy bins.

\section{Conclusions} \label{results}
The atmospheric $\nu_\mu+\overline \nu_\mu$ energy spectrum averaged over the upgoing hemisphere has been measured with the ANTARES neutrino telescope from 100 GeV to 200 TeV. 
Two different procedures based on different muon energy estimators have been used to unfold the neutrino spectrum. 
This measurement uses sea water as detection medium, which has completely different systematic uncertainties with respect to the stratified ice of the Antarctic. 

\begin{figure}
  \centering
  \includegraphics[width=1.1\textwidth]{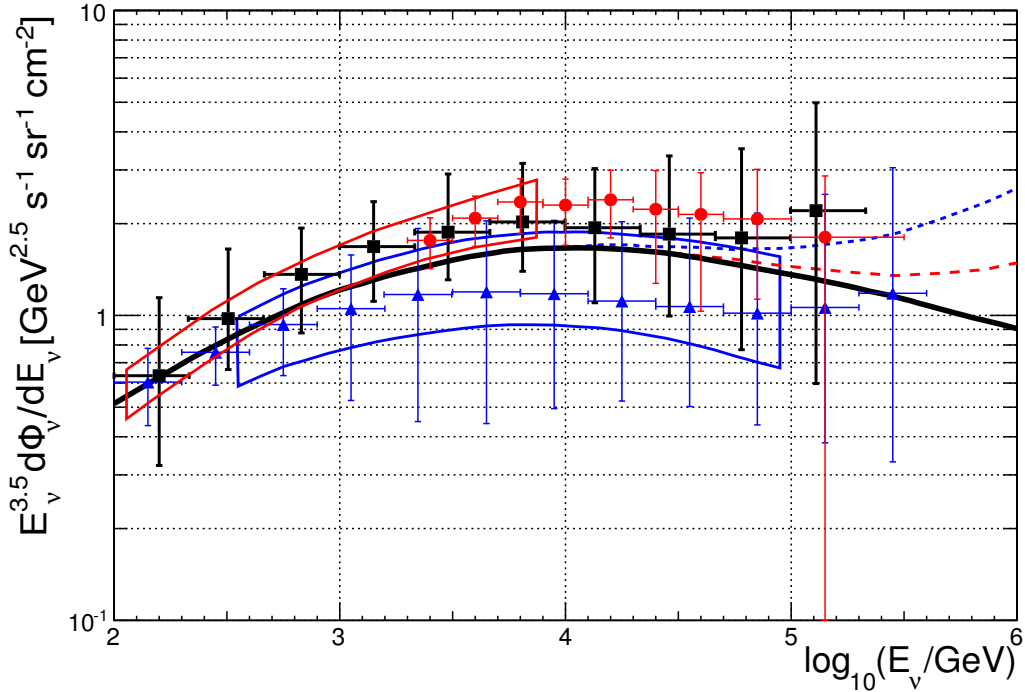}
  \caption{The atmospheric neutrino energy spectrum $E_\nu^{3.5} d\Phi_\nu/d E_\nu$ measured in this work in the zenith angle region $\theta>90^\circ$ (black full squares).
The full line represents the $\nu_\mu$ flux from Ref. \cite{barr}.  The red and blue dashed lines include two prompt neutrino production models from Ref. \cite{martin} and Ref. \cite{enberg}, respectively. All theoretical expectations are zenith-averaged from 90$^\circ$ to 180$^\circ$. 
The result of the AMANDA-II unfolding \cite{amanda_atm} averaged in the region 100$^\circ$ to 180$^\circ$ is shown with red circles and that of IceCube40 \cite{ic_atm} zenith-averaged from 97$^\circ$ to 180$^\circ$ is shown with blue triangles. The red region corresponds to the $\nu_\mu$ measurement from Ref. \cite{amanda_ff}, and the blue one the IC40 update from Ref. \cite{ic_diff}.}
  \label{fig:SpectrumExp}
\end{figure}

Figure \ref{fig:SpectrumExp} shows the result of the present measurement, where the atmospheric $\nu_\mu$ energy spectrum is averaged over zenith angle from 90$^\circ$ to 180$^\circ$.
The black line represents the conventional Bartol neutrino flux. The decreases above $E_\nu\sim 100$ TeV is expected from the change of the spectral index $\gamma_p$ of the primary cosmic ray flux above the \textit{knee} region ($E_p \ge 3\times 10^{15}$ eV).  

For comparison, the results from the Antarctic neutrino telescopes AMANDA-II \cite{amanda_atm} and IceCube40 \cite{ic_atm} are also shown. These two measurements average the zenith angle flux from 100$^\circ$ to 180$^\circ$ and 97$^\circ$ to 180$^\circ$, respectively.
 Assuming the expected angular distribution from the Bartol theoretical model, the flux integrated in the region $\theta> 90 ^\circ$ is larger than that obtained for $\theta> 100^\circ$ by factors of $\sim$ 3\%, 8\%, 25\% and 40\% at 0.1, 1, 10 and 100 TeV, respectively. The energy spectrum measured by ANTARES has a spectral index parameter $\gamma_{meas}=3.58\pm 0.12$ and the overall normalisation is 25\% larger than expected in Ref. \cite{barr}, almost uniformly in the measured energy range. 
This larger normalisation is also compatible with measurements from the MACRO underground experiment \cite{macro_fin}.

As in the case of Antarctic experiments \cite{amanda_atm,ic_atm}, the presence of a prompt contribution to the neutrino flux has not been established, even if some extreme contribution from prompt neutrino models have already been ruled out \cite{ic_diff}.

\vskip0.3cm
\noindent \textbf{Acknowledgements}
The authors acknowledge the financial support of the funding agencies: Centre National de la Recherche Scientifique (CNRS), Commissariat \'a l'\'energie atomique et aux energies alternatives  (CEA), Agence National de la Recherche (ANR), Commission Europ\'enne (FEDER fund and Marie Curie Program), R\'egion Alsace (contrat CPER), R\'egion Provence-Alpes-C\^ote d'Azur, D\'epartement du Var and Ville de La Seyne-sur-Mer, France; Bundesministerium f\"ur Bildung und Forschung (BMBF), Germany; Istituto Nazionale di Fisica Nucleare (INFN), Italy; Stichting voor Fundamenteel Onderzoek der Materie (FOM), Nederlandse organisatie voor Wetenschappelijk Onderzoek (NWO), The Netherlands; Council of the President of the Russian Federation for young scientists and leading scientific schools supporting grants, Russia; National Authority for Scientific Research (ANCS-UEFISCDI), Romania; Ministerio de Ciencia e Innovación (MICINN), Prometeo of Generalitat Valenciana  and MultiDark, Spain; Agence de l'Oriental, Morocco. Technical support of Ifremer, AIM and Foselev Marine for the sea operation and the CC-IN2P3 for the computing facilities is acknowledged.

\vskip0.3cm

\end{document}